\documentclass[12pt,twoside,fleqn]{article}
\usepackage{epsfig}
\usepackage{psfig}
\def\lsim{\raise0.3ex\hbox{$<$\kern-0.75em\raise-1.1ex\hbox{$\sim$}}}
\def\gsim{\raise0.3ex\hbox{$>$\kern-0.75em\raise-1.1ex\hbox{$\sim$}}}
\makeatletter
\@addtoreset{equation}{section}
\makeatother

\newcommand{\beqn} {\begin{equation}}
\newcommand{\eqn} {\end{equation}}

\newcommand{\slsh}[1] {#1\kern-.43em/}
\newcommand{\real}{{\sf I}\kern-.12em{\sf R}}
\newcommand{\comp}{{\sf I}\kern-.48em{\sf C}}
\newcommand{\nin} {\in\kern-.6em/}
\newcommand{\Tr} {\mbox{Tr}}

\def\MEF{m_{\rm eff}}\def\mef{\ifmmode\MEF\else$\MEF$\fi}

\def\SM{s_{\mu}}\def\xm{\ifmmode\SM\else$\SM$\fi}

\begin{document}
\thispagestyle{empty}
%
 \mbox{} \hfill BI-TP 96/24\\
 \mbox{} \hfill August 1996\\
\begin{center}
{{\large \bf SU(3) Latent Heat and Surface Tension from} \\
\bigskip
{\large \bf Tree Level and Tadpole Improved Actions}
} \\
\vspace*{1.0cm}
{\large B. Beinlich, F. Karsch and A. Peikert} \\
\vspace*{1.0cm}
{\normalsize
$\mbox{}$ {Fakult\"at f\"ur Physik, Universit\"at Bielefeld,
D-33615 Bielefeld, Germany}
}
\end{center}
\vspace*{1.0cm}
\centerline{\large ABSTRACT}

\baselineskip 20pt

\noindent
We analyze the latent heat and surface tension at the SU(3)
deconfinement phase transition with tree level and tadpole improved
Symanzik actions on lattices with temporal extent $N_\tau = 3$ and 4 and 
spatial extent $N_\sigma/ N_\tau = 4$, 6 and 8. In comparison to the
standard Wilson action we do find a drastic reduction of cut-off effects 
already with tree level improved actions. On lattices with temporal extent
$N_\tau=4$ results for the surface tension and latent heat obtained with a
tree level improved action agree well with those obtained with a tadpole 
improved action. A comparison with $N_\tau=3$ calculations, however, shows
that results obtained with tadpole action remain unaffected by cut-off effects 
even on this coarse lattice, while the tree level action becomes sensitive
to the cut-off.

\noindent
For the surface tension and latent heat we 
find $\sigma_I/ T_c^3 = 0.0155~(16)$ and $\Delta\epsilon/T_c^4 = 1.40~(9)$,
respectively.  

\vskip 20pt
\vfill
\eject
\baselineskip 15pt

\section{Introduction}

Tree level improved Symanzik actions \cite{Sym83} have been shown to
yield a large
reduction  of finite cut-off effects on bulk thermodynamic observables
even at temperatures close to $T_c$ \cite{Bei96,Bei96b}. It has been found 
that the large cut-off dependence observed in calculations with the standard
Wilson action \cite{Boy96} to a large extent does result from cut-off 
dependent corrections to the high temperature ideal gas limit. They,
therefore, can be
drastically reduced already with tree level improved Symanzik actions
\cite{Bei96,Bei96b} or fixed point actions \cite{perf}, which are constructed 
in order to reduce or eliminate the cut-off dependence in the weak coupling
/ high temperature
limit. Further cut-off dependent corrections of ${\cal O}(a^2g^2)$ do seem to 
be small for thermodynamic observables in the high temperature plasma phase. 
This can be seen, for instance, from a calculation of the difference of 
energy density and three times 
the pressure. The leading ideal gas contributions cancel in this quantity.
It thus is sensitive only to ${\cal O} (g^2)$ and higher order corrections.
Calculations of this quantity performed with the tree level improved
Symanzik action on lattices with small temporal extent 
do agree with the continuum extrapolation obtained from
simulations with the Wilson action \cite{Bei96b}. 
Bulk thermodynamics of $SU(N)$ gauge theories can thus be studied with
tree level improved actions showing 
little cut-off distortion already on lattices with temporal extent
$N_\tau =4$.    

The success of improved actions for the calculation of bulk
thermodynamics even at temperatures close to $T_c$ naturally leads to
the question whether these actions also do lead to an improvement at
$T_c$. This is, of course, a highly non-perturbative regime. However,
observables like the latent heat and the surface tension, which
characterize the discontinuities at the first order deconfinement phase 
transition in a
$SU(3)$ gauge theory, do depend on properties of the low as well as the
high temperature phase. As the latter is largely controlled by high
momentum modes it may be expected that some improvement does result
already from a perturbatively improved description of the high momentum 
modes in the deconfined phase. We will show here that this is indeed the case.
In addition we will address the question to what extent a tadpole improved 
Symanzik action \cite{Lep93,Bli96} does lead to a further reduction of the cut-off
dependence and would allow to perform calculations on even coarser lattices. 

The latent heat and the surface tension at the
SU(3) deconfinement transition have been studied with the standard Wilson
action on lattices up to
temporal extent $N_\tau=6$ \cite{Iwa92,Iwa94}. A strong cut-off 
dependence has been found when comparing calculations for $N_\tau =4$ 
and 6 which is compatible in magnitude with
the cut-off dependence visible in calculations of the energy density or
pressure in the plasma phase. On the basis of these
results an extrapolation to the continuum limit for these observables is
not yet possible and calculations on larger temporal lattices do not seem to
be feasible with the standard Wilson action given the present computational
possibilities. One thus has to aim at a reduction of the cut-off
dependence on lattices with small temporal extent in order to get
reliable results for the continuum values of the latent heat and surface
tension at $T_c$. We will present here results from calculations with
a tree level as well as a tadpole
improved Symanzik action on lattices of temporal extent 
$N_\tau =3$ and 4 and spatial extent $N_\sigma / N_\tau$ ranging from 4 to 8.

\section{Critical Couplings}

We use the (1,2) Symanzik action which is defined by adding a planar (1,2) 
Wilson loop to the standard Wilson action defined in terms of a
(1,1) Wilson loop (plaquette),
\begin{eqnarray}
S (\beta) & = &   \sum_{x, \nu > \mu} \biggl( \frac{5}{3}
W^{1,1}_{\mu, \nu}(x)  - \frac{1}{6 u_0^2(\beta)}
W^{1,2}_{\mu, \nu}(x) \biggr) 
~~.
\label{tree}
\end{eqnarray}
Here $W^{k,l}_{\mu, \nu}$ denotes a symmetrized combination of $k\times l$
Wilson loops in the $(\mu, \nu)$-plane of the lattice,
\beqn
W^{k,l}_{\mu, \nu} (x) = 1- {1 \over 2N} \biggl( {\rm Re~Tr} V^{(k)}_{x,\mu}
V^{(l)}_{x+k\hat\mu,\nu} V^{(k)+}_{x+l\hat\nu,\mu} V^{(l)+}_{x,\nu}
+ (k \leftrightarrow l) \biggr)~~,
\label{wloop}
\eqn
with
$ V^{(k)}_{x,\mu} =\prod_{j=0}^{k-1} U_{x+j\hat\mu,\mu}$ and
$x=(n_1,n_2,n_3,n_4)$ denoting the sites on an
asymmetric lattice of size $N_\sigma^3 N_\tau$.
We consider the tree level improved action, corresponding to $u_0\equiv 1$,
as well as a tadpole improved action \cite{Lep93} where 

\beqn
u_0^4 = {1 \over 6 N_\sigma^3 N_\tau}~
\langle \sum_{x, \nu > \mu} (1-W^{1,1}_{\mu, \nu} (x))~\rangle
\label{u0}
\eqn
is the self-consistently determined plaquette expectation
value. We have determined this tadpole improvement factor at several
values of the gauge coupling $\beta$ and then used a spline interpolation in
order to define $u_0$ at intermediate values of $\beta$. This
allows to give an unambiguous definition of derivatives of the tadpole 
term with respect
to the gauge coupling which is needed in thermodynamic observables.
Moreover, it allows us to use the Ferrenberg-Swendsen reweighting technique in
the vicinity of $T_c$. We note that the tadpole improved action does
depend on the gauge coupling $\beta$ through $u_0$. This does lead to
some modifications of the relations usually explored for thermodynamic
observables, which are defined through derivatives of the partition
function, $Z = \int {\cal D}U \exp(-\beta S(\beta))$, with respect to
$\beta$. For the purpose of the calculations presented here this is of 
relevance for the analysis of the latent heat. 

Most of our calculations have been performed close to the critical coupling on
lattices with temporal extent $N_\tau=3$ and 4. On 
the smaller lattices we
have performed simulations at several nearby $\beta$-values. 
Typically 20.000-40.000 configurations have been generated 
at each $\beta$-value using an overrelaxed heat bath algorithm (1 iteration = 
4 overrelaxation + 1 heat bath step). The critical 
couplings have then been determined from the location of peaks in the 
Polyakov-loop susceptibilities, 
\beqn
\chi = \langle |L|^2\rangle - \langle |L| \rangle^2 ~~,
\label{chi}
\eqn
where $L$ denotes the Polyakov-loop
\beqn
L = {1 \over N_\sigma^3} \sum_{\vec{x}} \Tr \prod_{x_4=1}^{N_\tau}
U_{(\vec{x},x_4),4}~~.
\label{polyakovloop}
\eqn
We have used an 
interpolation based on the Ferrenberg-Swendsen reweighting method to
determine the location of the maximum in $\chi$.  
In Table~\ref{tab:betac} we summarize our results for the critical
couplings. In the case of the Symanzik improved tree level action our results
do agree well with those of earlier calculations \cite{Cel94}. 
For $N_\tau=4$ we have used results from different spatial
lattice sizes to perform an infinite volume extrapolation of the critical
couplings using the ansatz 
\beqn
\beta_{c}(N_\tau) = \beta_{c}(\infty) - h (N_\tau / N_\sigma)^3 
\label{betac}
\eqn
which is valid for first order phase transitions.
The coefficient $h$ characterizing the infrared sensitivity of the
thermodynamics at $T_c$ has been found to be similar for both improved actions. 
We find $h=0.101~(34)$ and 0.068~(45) for the tree level and tadpole
improved actions, respectively. This also agrees well with the volume
dependence observed in the case of the standard Wilson action \cite{Iwa92,Fuk90}.

\begin{table}[hbt]
\catcode`?=\active \def?{\kern\digitwidth}

\vskip 5pt
\begin{center}
\begin{tabular}{|c|c|c|r|c|c|r|}\hline
&\multicolumn{3}{|c|}{tree level}&\multicolumn{3}{|c|}{tadpole}\\ \hline
$N_\sigma^3 \times N_\tau$& $\beta_c$& \#$\beta$ &\# iterations&
$\beta_c$& \#$\beta$ &\# iterations \\ \hline
\hline
$12^3\times 3$&3.9079~(6)& 5 & 80000 & 4.1868~(4)& 6 & 85500 \\
\hline
\hline
$16^3\times 4$&4.0715~(4)& 8 & 157500 & 4.3512~(4)& 7 & 133400 \\
$24^3\times 4$&4.0722~(4)& 3 & 75200 & 4.3519~(5)& 1 & 38500 \\
$32^3\times 4$&4.0729~(3)& 1 & 46800 & 4.3522~(4)& 1 & 59000 \\
\hline
\hline
$(\infty)^3 \times 4$&4.0730~(3)& - & - & 4.3523~(4)& - & - \\
\hline
\end{tabular}
\end{center}
\caption{Critical couplings for the tree level and tadpole improved
actions. Errors are obtained from a jack-knife analysis. Columns four 
and seven give the combined statistics from 
different $\beta$-values entering
the Ferrenberg-Swendsen analysis of the susceptibilities, surface tension
and latent heat. The number of $\beta$-values entering this analysis are
given in columns three and six.} 
\label{tab:betac}

\end{table}

\section{Surface Tension}

The $SU(3)$ deconfinement transition is known to be of first order.
At the critical temperature the deconfined and confined phases can
coexist in a mixed phase. Both coexisting phases are separated by
an interface. The additional energy needed to build up such interfaces
does, however, make these configurations more unlikely than pure states
of one or the other phase. This is reflected in the
double-peaked probability distribution functions of observables at the
critical point. While the two maxima are proportional to the probability
of finding the system in one of the two pure states the
minimum inbetween is related to the probability of a state containing
two phases separated by an interface.

To be specific we consider the probability distribution of the absolute
value of the Polyakov loop, $|L|$,
and follow the analysis presented in Ref.~\cite{Iwa94} to extract
the surface tension. The probability distribution at the minimum is
proportional to
\beqn
P(|L|) \sim \exp\bigl(- \bigl[f_1 V_1 +f_2 V_2 + 2 \sigma_I A\bigr]/T \bigr)
\label{prob}
\eqn
where $f_i$ denotes the free energy in the phase $i$, and $V_i$ is the
volume occupied by that phase. The last term denotes the extra free
energy needed to create a surface to separate both phases. Here $\sigma_I$
denotes the surface free energy and $A$ is the area of the interface. An
additional factor of two does appear here because we consider finite volumes
with periodic boundary conditions and one always has to create
two interfaces in this case. At
$T_c$ the free energy density in both phases is identical, $f_1=f_2$. 
In our simulations we have observed frequent flips between the
confined and deconfined phase on the $12^3,~16^3$ and $24^3$ lattices
which suggests that we have properly sampled both phases.
On the $32^3$ lattice the data samples include, however,
only 4 - 5 flips at each $\beta$-value. 

In order to determine the maxima and the minimum of $P(|L|)$ 
the region in the vicinity of the extrema has been fitted 
with third order polynomials \cite{Iwa94}. 
A Ferrenberg-Swendsen reweighting 
has been used to shift the $\beta$-value to a point where the two
maxima, $P_{max,1}$ and $P_{max,2}$, in the distribution function of the 
Polyakov-loop have equal height.
\begin{figure}[htb]
\centerline{
   \epsfig{
       file=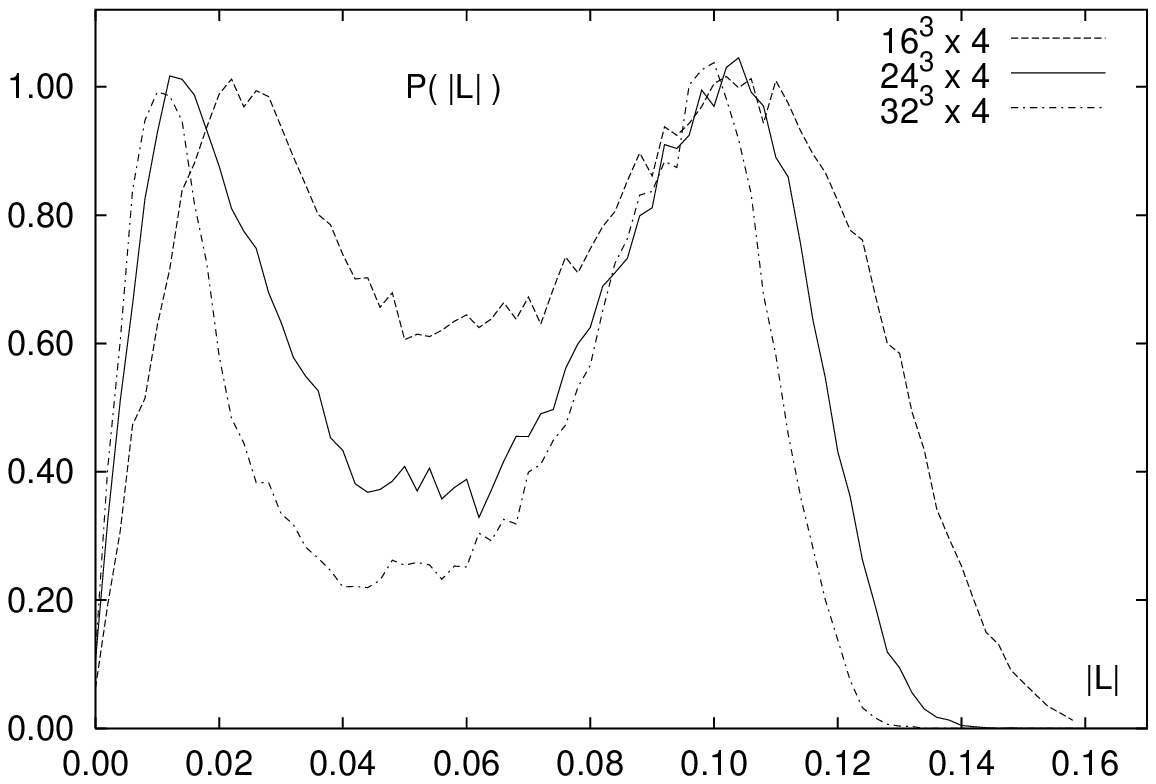,width=90mm} }
\vskip 1.0truecm
\centerline{
   \epsfig{
       file=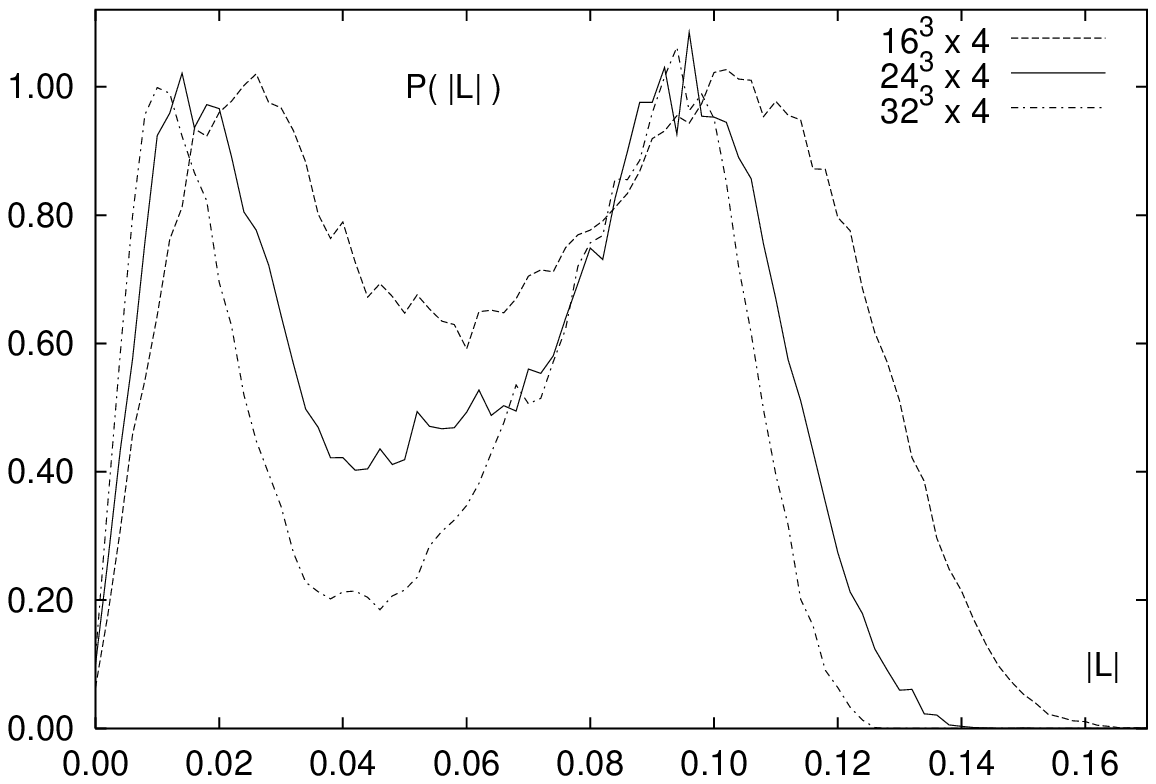,width=90mm}}
\caption{\baselineskip 10pt
Probability distributions of $|L|$ at the critical couplings.
}
\label{fig:surface}
\end{figure}
A simple estimate of the surface tension is then given by 
\beqn
\biggl({\sigma_I \over T_c^3}\biggr)_V = - 
\frac{1}{2}\biggl({N_\tau \over N_\sigma} \biggr)^2 \ln(P_{min}/P_{max}) ~~,
\label{surface}
\eqn
where $P_{max}\equiv P_{max,1} \equiv P_{max,2}$. In order to obtain a
reliable error estimate we have performed a jack-knife analysis
at this $\beta$-value. As the relative maxima are then no longer of equal
height in the course of the analysis, we define the surface tension through
the jack-knife average of
\beqn
\biggl({\sigma_I \over T_c^3}\biggr)_V = - 
\frac{1}{2}\biggl({N_\tau \over N_\sigma} \biggr)^2 
\ln\biggl( {P_{min} \over P_{max,1}^{\gamma_1} P_{max,2}^{\gamma_2}} \biggr) ~~.
\label{surfacejack}
\eqn
where $\gamma_1$ and $\gamma_2$ characterize the relative weight of the
confined and deconfined  
phases at $\beta_c$ and are fixed by demanding that the Polyakov loop
expectation value is given by the weighted sum of the values $L_1$ and
$L_2$ at $P_{max,1}$ and $P_{max,2}$, respectively, i.e.
\beqn
\langle |L| \rangle = \gamma_1 L_1 +  \gamma_2 L_2 \quad {\rm with}\quad
\gamma_1 + \gamma_2 = 1~~.
\label{Ljack}
\eqn
Of course, we do find that the surface tension defined this way agrees
within errors with the global value extracted from Eq.~\ref{surface}. For 
the relative weight $\gamma_1$ we find in all cases values close to 0.4
which also is similar to the case of the standard Wilson action\footnote{There
is a misprint in the labeling of the last column of Table I in 
Ref.~\cite{Iwa94}. This column does give $\gamma_2$ rather than $\gamma_1$.}
\cite{Iwa94}. 

The results for $\sigma_I/ T_c^3$ are given in Table~\ref{tab:surface} 
for the improved actions
analyzed by us on lattices with temporal extent $N_\tau = 3$ and 4. Also
given there are the results from \cite{Iwa94}
obtained with the standard Wilson action on $N_\tau = 4$ and $N_\tau =
6$ lattices. 

\begin{table}[hbt]
\catcode`?=\active \def?{\kern\digitwidth}

\vskip 5pt
\begin{center}
\begin{tabular}{|c|c|c|c|c|}\hline
&&\multicolumn{3}{c|}{$\sigma_I/T_c^3$}\\ \hline
$N_\tau$&volume& tree level&tadpole&Wilson\\ 
\hline \hline
3&$12^3$ & 0.0234~(24)& 0.0158~(11) &\\
\hline \hline
4&$12^2\times 24$&&&0.0241~(27)\\
4&$24^2\times 36$&&&0.0300~(16)\\ \hline
4&$16^3$&0.0148~(16)&0.0147~(14)&\\
4&$24^3$&0.0136~(25)&0.0119~(21)&\\ 
4&$32^3$&0.0116~(23)&0.0125~(17)&\\ \hline \hline
4&$\infty$ &0.0152~(26)&0.0152~(20)&0.0295~(21) \\ \hline\hline
6&$20^3$&&&0.0123~(28)\\
6&$24^3$&&&0.0143~(22)\\
6&$36^2\times 48$&&&0.0164~(26)\\ \hline \hline
6&$\infty$ & -- & -- &0.0218~(33) \\
\hline
\hline
\end{tabular}
\end{center}
\caption{Surface tension for the improved actions 
on several finite spatial lattices with temporal extent $N_\tau =4$ and
the Wilson action for $N_\tau =4$ and 6.  The latter are taken from Ref.~[9].
The second column gives the spatial lattice size.  }
\label{tab:surface}

\end{table}

In order to extract the infinite volume result for the surface tension one
has to take into account finite volume corrections, which result from
the finite width of the (gaussian) peaks in the pure phases, the
contribution of zero modes resulting from the translational invariance as
well as from contributions of fluctuations of the interface.
Taking these into account, the interface tension should be extrapolated to
infinite volume using the ansatz given in Ref.~\cite{Iwa94}, 
\beqn
\biggl({\sigma_I \over T_c^3}\biggr)_V = 
\biggl({\sigma_I \over T_c^3}\biggr) - \biggl({N_\tau \over N_\sigma}
\biggr)^2 \biggl[ c + {1\over 4} \ln N_\sigma \biggr] ~~.
\label{surffit}
\eqn
The result of such an extrapolation is also given in
Table~\ref{tab:surface} for the $N_\tau =4$ calculations.
Even without performing the infinite volume extrapolations, it is,
however, obvious already from Table~\ref{tab:surface} that the surface
tension is reduced in the improved action calculations relative to
that of the Wilson action with the same temporal extent.

We note that the tree level improved action still shows 
some cut-off dependence when comparing the $N_\tau=3$ and 4 results while the
results for the tadpole improved action coincide within errors. Combining
both results for the tadpole improved action we find as an estimate for the 
surface tension 
\beqn
{\sigma_I \over T_c^3} = 0.0155 \pm 0.0016~~.
\label{surfresult}
\eqn
It is interesting to note that this agrees well with an extrapolation in
$1/N_\tau^2 = (aT_c)^2$ of the results obtained with the Wilson action for 
$N_\tau = 4$ and 6.

\section{The Latent Heat}

The latent heat is calculated from the discontinuity in the energy
density, $\epsilon$, or more conveniently directly from the
discontinuity in ($\epsilon -3P)$. The latter is obtained from the
discontinuity in the various Wilson loops entering the definition of
the improved actions,
\begin{eqnarray}
{\Delta\epsilon \over T_c^4} &=& \Delta\biggl( {\epsilon - 3P \over
T_c^4} \biggr) \quad = \quad {1\over 6}
\biggl({N_\tau \over N_\sigma} \biggr)^3 \biggl( a{{\rm d}\beta \over {\rm d} a}
\biggr) \biggl(\langle \tilde{S} \rangle_+  - \langle \tilde{S} \rangle_-
\biggr) ~~,
\label{latent}
\end{eqnarray}
with
$
\tilde{S}  \equiv  S - {{\rm d} S / {\rm d} \beta} 
$
and $a{{\rm d}\beta / {\rm d} a}$ denoting the SU(3) $\beta$-function. 

The difference of the expectation values at $\beta_c$ is obtained by 
calculating expectation values separately in the two coexisting phases at the
critical coupling $\beta_c$. This does require large spatial lattices in
order to separate well the two coexisting phases. 
We therefore have performed the analysis of the latent heat only for the 
$32^3\times 4$ lattices. Following the approach
used in Ref.~\cite{Iwa92} we have analyzed the time histories of the 
Polyakov loop values and introduced a cut to classify configurations 
belonging to either of the two phases. We then have performed averages in 
the two phases separately and left out configurations belonging to the 
transition region. We have checked that the expectation values calculated 
in both phases are within errors insensitive to the precise 
choice of the cut and the width of transition regions.
\begin{figure}[htb]
\centerline{
   \epsfig{
       file=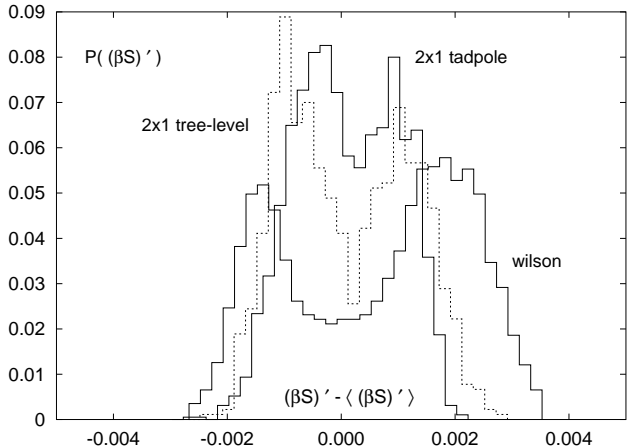,width=90mm}}
\caption{\baselineskip 10pt
Probability distributions of $\tilde{S}$ at the critical couplings.
}
\label{fig:latent}
\end{figure}
In Figure~\ref{fig:latent} we show the distribution of $\tilde{S}$ for
the improved actions as well as the Wilson action on lattices
with temporal extent $N_\tau = 4$. It is obvious from this Figure that
the discontinuity is drastically reduced for the improved actions
relative to the Wilson action. 

In order to extract the latent heat we
still need the $\beta$-function entering the definition of $\Delta
\epsilon /T_c^4$ in Eq.~\ref{latent}. We have determined this through
the calculation of the string tension in units of the cut-off
following the approach discussed in Ref.~\cite{Boy96}. These calculations
have been performed on lattices of size $16^4$ \cite{Bei96b}. 
We also have reanalyzed the results for the latent heat presented in
\cite{Iwa92} using the newly determined $\beta$-function of \cite{Boy96}.
Further details of our determination of the
temperature scale will be presented in Ref.~\cite{Bei96b}. 

The resulting
values for the latent heat are summarized in Table ~\ref{tab:latent}.
\begin{table}[hbt]
\catcode`?=\active \def?{\kern\digitwidth}

\vskip 5pt
\begin{center}
\begin{tabular}{|l|c|c|c|c|}\hline
action&$V_\sigma$&$N_\tau$&$\Delta\epsilon/T_c^4$ &
$\biggl(\Delta\epsilon/T_c^4\biggr)_{\rm pert}$\\ \hline
standard Wilson &$24^2\times 36$&4&2.27~(5) &4.06~(8)~\\
&$36^2\times 48$&6&1.53~(4) &2.39~(6)~\\ \hline
(1,2) Symanzik (tree level) &$32^3$&4&~1.57~(12) &~2.28~(8)~\\
(1,2) Symanzik (tadpole) &$32^3$&4&~1.40~(9) &~1.94~(8)~\\ \hline
\end{tabular}
\end{center}
\caption{Latent heat for the three improved actions and the Wilson
action. Results for the Wilson action are taken from [8] using
the non-perturbative $\beta$-function calculated in Ref.~[4].
}
\label{tab:latent}
\end{table}
There we also give results obtained with the perturbative $\beta$-function,
$a{\rm d}\beta /{\rm d}a  = 33/4\pi^2$. The ratio of the fourth and fifth 
column in Table~\ref{tab:latent} thus reflects the deviations of the
non-perturbative determination of the $\beta$-functions from the
perturbative result. We note that these ratios are, in fact, quite
similar, which shows that the improved actions do not lead to an
improved asymptotic scaling behaviour. The results summarized in
Table~\ref{tab:latent} do, however, show that the improved actions do
reduce drastically the cut-off dependence of the latent heat. Results
obtained with the tree level and tadpole improved actions for $N_\tau =4$ agree
with each other. They are about
30\% smaller than in the case of the Wilson action and are compatible
with the $N_\tau=6$ results for this action. As a best estimate for the
latent heat we obtain
\beqn
{\Delta \epsilon \over T_c^4} = 1.40 \pm 0.09~~.
\label{latentest}
\eqn

\section{Conclusion}

We have analyzed the surface tension and the latent heat at the first order
deconfinement phase transition of the SU(3) gauge theory using tree level
and tadpole improved actions. We do find that these actions drastically
reduce the cut-off dependence observed previously in calculations of 
these quantities with the standard Wilson action. Results obtained with a
Symanzik improved tree level action agree within errors with those obtained 
with a tadpole improved action for $N_\tau =4$ lattices. The results
obtained with the tree level action do, however, show a clear cut-off
dependence when comparing  
calculations on the $N_\tau =3$ and 4 lattices while results
obtained with the tadpole improved action remain unaffected by the change in
lattice cut-off. 
We thus expect that our results obtained on the $N_\tau = 4$ lattice with 
the tadpole improved action do provide a good estimate for the continuum
limit result of discontinuities in thermodynamic observables at $T_c$. 

Over the small range of couplings we have explored in this study the tadpole 
improvement factor varies only slightly, $0.86 < u_0(\beta) < 0.88$.
Although we do take into account this variation in our simulations we note
that we do effectively
work with an {\it over-improved} Symanzik action where the weight of the
contribution of $(1\times 2)$-loops relative to the $(1\times 1)$-plaquette
term is increased from $1/10$ by about 30\%. Testing the role of the
$\beta$-dependence of the tadpole improvement factor does require an
analysis of the thermodynamics in a larger temperature interval. This will
be presented elsewhere \cite{Bei96b}.  It also would be interesting to
investigate the sensitivity of the analysis presented here to even larger
changes in the weight of the $(1\times 2)$-loop contribution. In the case
of the renormalization group improved action \cite{Ito86,Iwa96}, for
instance, the relative weight is as large as 0.1815. 

\vskip 2pt
\noindent
{\large \bf Acknowledgements:}
\vskip -7pt

\noindent
The work has been supported by the DFG under grant Pe 340/3-3.
Numerical calculations have been performed on the QUADRICS parallel 
computers funded by the DFG under contract no. Pe 340/6-1.


%
\end{document}